# Expanding the Horizons of Phase Transition-Based Luminescence Thermometry

M. Tahir Abbas[1], M. Szymczak[1], V. Kinzhybalo[1], D. Szymanski[1], M. Drozd[1], L. Marciniak[1*]

[1] Institute of Low Temperature and Structure Research, Polish Academy of Sciences, Okólna 2, 50-422 Wrocław, Poland

*corresponding author: l.marciniak@intibs.pl

**Abstract**

The limited operational range of phase transition-based luminescence thermometers necessitates the exploration of new host materials exhibiting first-order structural phase transitions to broaden the applicability of this approach. Addressing this need, the present study investigates the spectroscopic properties of $LaGaO_3$:$Eu^{3+}$ as a function of temperature. A thermally induced structural transition from the low-temperature orthorhombic phase to the high-temperature trigonal phase, occurring at approximately 430 K, significantly alters the spectroscopic properties of $Eu^{3+}$ ions. Specifically, a reduction in the number of Stark lines due to changes in the point symmetry of $Eu^{3+}$ ions enables the development of a ratiometric luminescence thermometer with sensitivity as high as $S_R$=6% $K^{-1}$ at 480 K. Furthermore, it was demonstrated that increasing the concentration of $Eu^{3+}$ ions shifts the phase transition temperature, allowing for modulation of the thermometric performance of this luminescence thermometer. The findings presented here not only expand the repertoire of phase transition-based luminescence thermometers but also illustrate how the luminescence properties of $Eu^{3+}$ ions can be employed to accurately monitor structural changes in the host material.

**Introduction**

The substantial increase in the number of scientific studies on luminescence thermometry over the past decade underscores the growing popularity and scientific as well as practical appeal of this field[1–5]. This trend can be attributed to the numerous advantages of this technique, including remote readout capability, electrically passive operation, and high measurement precision[3,6–8]. Moreover, the application of luminescence in temperature sensing has enabled the development of numerous novel applications that were previously unattainable [9–11]. The efficacy of luminescence thermometry has been well-documented in diverse fields such as in vivo bioimaging[12–15], theranostics[16–18], turbulent gas flow systems[19–23], catalytic processes[24–26], and beyond[27,28]. These investigations not only introduce novel thermometric materials but also propose innovative strategies to enhance the thermometric performance and address the stringent requirements of specific applications. A key advantage of luminescence thermometry lies in the tunability of the spectroscopic and thermometric properties of inorganic materials through chemical engineering[29]. This flexibility allows for the customization of spectral emission ranges, relative sensitivities, thermal operating ranges, and spectral excitation ranges to meet diverse application demands. Among the various materials employed for remote temperature sensing-such as diamonds[30,31], carbon dots[13,14], metal-organic frameworks (MOFs)[32–34], organic dyes[35], and inorganic phosphors doped with transition metal[29,36–39] or lanthanide ions[40–42]-the latter group has received the most research attention. This popularity is largely due to the unique energy level configurations of lanthanide ions, which, in certain cases, exhibit thermal coupling between two emitting multiplets[43–45]. The population ratios of these levels, and the resulting luminescence intensities from their radiative depopulation, can be described by the Boltzmann distribution within a specific temperature range[46]. This property is pivotal for the development of primary luminescence thermometers[47–50]. The relative sensitivity of these thermometers is proportional to the energy separation between the thermally coupled levels, with efficient coupling typically observed for energy gaps below ~2000 cm$^{-1}$. However, the primary limitation of such thermometers is their relatively low relative sensitivity, usually below 2% K$^{-1}$. Consequently, research efforts have focused on strategies to enhance sensitivity. One promising approach involves leveraging thermally induced structural changes in the host material, particularly those associated with first-order phase transitions[51–56]. In these cases, changes in the crystallographic symmetry of the host material during a phase transition alter the spectroscopic properties of the incorporated lanthanide ions.

While lanthanide ions are generally considered minimally sensitive to local environmental changes due to the predominance of intra-configurational *4f-4f* transitions, structural changes induced by phase transitions can influence the intensity of electric dipole-type transitions or the splitting of multiplets via the Stark effect[44,57–60]. These changes enable the development of highly sensitive ratiometric luminescence thermometers[51,55]. A notable example is LiYO$_2$, which undergoes a monoclinic-to-tetrahedral phase transition near room temperature[61–66]. By selecting appropriate Ln$^{3+}$ dopants, the operational spectral range of such thermometers can be extended across the visible and infrared regions, with Eu$^{3+}$ [53,67], Pr$^{3+}$ [68], Er$^{3+}$ [53], Nd$^{3+}$ [69], and Yb$^{3+}$ [70] ions being the most frequently reported. The rapid spectral emission changes near the phase transition temperature in these phosphors result in exceptionally high relative sensitivities, often exceeding 10% K$^{-1}$ [67] and, in some cases, reaching even 35% K$^{-1}$ [55]. However, the thermal operating range of such materials is constrained by the host material's structural transformation.

A partial solution to this limitation has been demonstrated through co-doping the host material with ions of varying ionic radii to shift the phase transition temperature[53]. However, this approach has been associated with a reduction in relative sensitivity. Thus, there remains a need to identify new host materials that exhibit phase transitions within a broader temperature range to expand their potential applications.

In response to these challenges, this study investigates $LaGaO_3$, which exhibits a structural phase transition around 460-570 K[71–76]. Upon exceeding this temperature, the host material transitions from a low-temperature orthorhombic structure to a high-temperature trigonal structure, altering the symmetry of $La^{3+}$ ions, which can be substituted with $Ln^{3+}$ dopant ions. The potential application of this phase transition in luminescence thermometry was assessed using $Eu^{3+}$ ions, known for their high sensitivity to changes in local crystallographic environments. The presented experiments were conducted as a function of temperature and $Eu^{3+}$ ion concentration, demonstrating the viability of $LaGaO_3$ as a host material for highly sensitive luminescence thermometers.

## 2. Experimental Section

*Materials*

The $LaGaO_3$: x%$Eu^{3+}$ (x = 0.1, 0.25, 0.5, 1, 2, 5) nanocrystals were synthesized using a modified Pechini method. $La_2O_3$ (99.999% purity, Stanford Materials Corporation), $Eu_2O_3$ (99.99% purity, Stanford Materials Corporation), $Ga(NO_3)_3 \cdot 9H_2O$ (99.999% purity, Alfa Aesar), $C_6H_8O_7$ (CA, >99.5% purity, Alfa Aesar) and $H(OCH_2CH_2)_nOH$, (PEG-200, n = 200, Alfa Aesar) were used as starting materials. A stoichiometric amounts of lanthanum(III) and europium(III) oxides were dissolved in distilled water with the addition of 3 ml of nitric acid (65% solution, Avantor). The solution was then recrystallized three times to remove excess of nitric acid. The $Ga(NO_3)_3 \cdot 9H_2O$ was added to the water solution of lanthanide nitrates after recrystallization. After that, citric acid and polyglycol were added to the mixture. The molar ratio of CA to all metal cations was set up as 6:1, meanwhile the PEG-200 was used in a 1:1 molar ratio relative to citric acid. The prepared solution was subsequently dried at 90 °C for three days until a resin was formed. The obtained resins were annealed in porcelain crucibles for 3 h in air at a temperature of 600 °C. After natural cooling to room temperature, the obtained samples were ground in an agate mortar and reheated at 1050 °C for 6 h in air atmosphere. Finally, the produced powders were ground thoroughly in an agate mortar for further characterization.

*Methods*

The X-ray powder diffraction measurements of the samples were carried out in Bragg-Brentano geometry using PANalytical X'Pert Pro diffractometer equipped with Anton Paar HTK 1200N high-temperature attachment using Ni-filtered Cu K$\alpha$ radiation ($V = 40$ kV, $I = 30$ mA). The measurements were performed in $10 - 90°$ $2\theta$ range. Variable temperature powder X-ray diffraction in the $298 - 623$ K range were carried out in $15 - 120°$ $2\theta$ range in the Bragg-Brentano geometry in the alumina sample holder. The excitation and emission spectra were recorded using the FLS1000 Fluorescence Spectrometer from Edinburgh Instruments equipped with 450 W Xenon lamp and R928 photomultiplier tube from Hamamatsu. To perform temperature – dependent measurements, the temperature of the sample was controlled by a THMS600 heating – cooling stage from Linkam (0.1 K temperature stability and 0.1 K point resolution). Luminescence decay profiles were also measured using FLS1000 equipped with 150 W µFlash lamp. The average lifetime ($\tau_{avr}$) of the excited states was determined by using of double – exponential function:

$$\tau_{avr} = \frac{A_1 \tau_1^2 + A_2 \tau_2^2}{A_1 \tau_1 + A_2 \tau_2} \qquad (1)$$

$$I(t) = I_0 + A_1 \cdot \exp\left(-\frac{t}{\tau_1}\right) + A_2 \cdot \exp\left(-\frac{t}{\tau_2}\right) \qquad (2)$$

where $\tau_1$ and $\tau_2$ are decay components and $A_1$ and $A_2$ are the amplitudes of the double – exponential function.

## 3. Results and discussion

The perovskite LaGaO$_3$ crystallizes in two structures: low-temperature orthorhombic phase with *Pnma* space group (No. 62, ICSD-182539) and high-temperature trigonal phase with $R\bar{3}c$ space group (No. 167, ICSD-182540))[71–76]. The LaGaO$_3$ has two different cation sites: the Ga$^{3+}$ site (B-site), which is coordinated with six oxygen atoms, forming a GaO$_6$ octahedron, and the La$^{3+}$ ion (A-site), which is surrounded by twelve oxygen atoms to form a dodecahedron (Figure 1a). The lanthanide ions occupy the 4c sites, possessing $C_s$ point symmetry in the orthorhombic phase *Pnma*[71–73]. In the high-temperature trigonal phase $R\bar{3}c$, the lanthanide ions are in the 6a sites, with point symmetry $D_3$. The LaGaO$_3$ undergoes the temperature-induced first order structural phase transition from orthorhombic to trigonal phase at about 418 K[74,75,77,78]. The phase transition leads to a change in the unit cell parameters from $a = 5.49$, $b = 7.77$, $c = 5.53$ Å to $a = 5.53$, $c = 13.40$ Å (hexagonal cell setting). Due to phase transition, the local point symmetry occupied by La$^{3+}$ cation changes from $C_s$ to $D_3$ with increasing temperature. Eu$^{3+}$ ions substitute La$^{3+}$ ions in LaGaO$_3$:Eu$^{3+}$ due to similar ionic radii and charge matching. The room temperature XRD patterns of LaGaO$_3$:x%Eu$^{3+}$ (x = 0.1, 0.25, 0.5, 1, 2) were compared with references, indicating that all synthesized materials are phase – pure, with no additional reflections detected (Figure 1b). The XRD patterns of LaGaO$_3$:0.25%Eu$^{3+}$ measured as a function of temperature (Figure S1-S18) revealed that at elevated temperature above around 400 K the reflections associated with the trigonal phase of LaGaO$_3$ starts to be observed. The Rietveld refinement of these patterns indicated that the contribution of the HT phase starts to rapidly increases above 400K reaching 50% around 430K (Figure 1c). Above 460 K no evidence of LT phase can be found in the XRD patterns. This observation clearly confirms the presence of the thermally induced phase transition in the synthesized LaGaO3:$Eu^{3+}$ powders. The DSC studies revealed that the temperature of this phase transition increases with $Eu^{3+}$ dopant concentration from 428 K for 0.1%$Eu^{3+}$ to 511 K for 2%$Eu^{3+}$ (Figure 1d). This effect associated with the difference in the ionic radii between host material cation ($La^{3+}$) and dopant ions was previously described in the literature[51]. Morphological studies confirm that the synthesized powders consist of microsized grains (Figure 1e) with the uniformly distributed elements (Figure 1f-h).

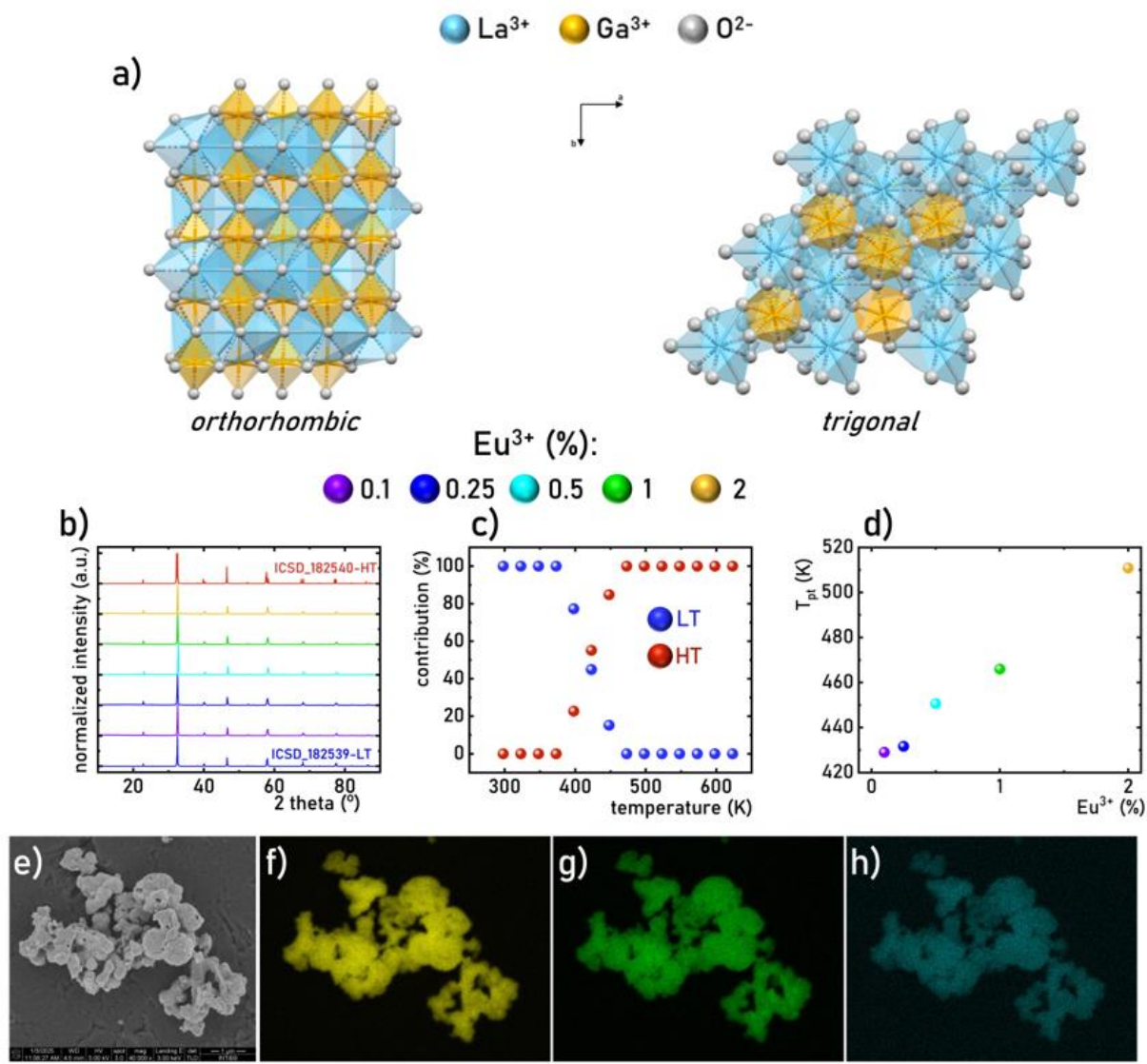

**Figure 1.** Visualization of the structure of low-temperature (LT) orthorhombic and high-temperature (HT) trigonal structures of LaGaO$_3$-a); room temperature XRD patterns for LaGaO$_3$:Eu$^{3+}$ with different concentrations of dopant ions-b); the influence of the temperature on the contribution of LT and HT phases of LaGaO$_3$ determined based on Rietveld refinement for LaGaO$_3$: 0.25%Eu$^{3+}$-c); the influence of Eu$^{3+}$ concentrations on the phase transition temperature -d); representative SEM image of LaGaO$_3$:0.25%Eu$^{3+}$-e) and corresponding elemental maps of the La (yellow), Ga (green) and Eu (cyan).

Lanthanide ion multiplets undergo splitting into Stark levels due to interactions with the electric field generated by the surrounding ions in the host material[57]. The number of Stark levels depends on the J quantum number of the multiplet and the local symmetry of the crystallographic site occupied by the ion[57]. Consequently, structural phase transitions that alter the local symmetry of a lanthanide ion are expected to manifest as changes in the number of Stark components within the emission spectrum. This property makes lanthanide ions particularly suitable for phase transition-based luminescence thermometry. Among lanthanide ions, $Eu^{3+}$ is an excellent candidate due to the non-degenerate nature of its $^5D_0$ emitting level, which does not split into Stark lines (Figure 2a) [57,79,80]. This ensures that changes in the emission band shape are driven solely by the redistribution of Stark levels in the levels receiving transitions from $^5D_0$, minimizing the risk of spectral overlap between Stark lines from different crystallographic phases. Such overlap could otherwise degrade the relative sensitivity of the luminescence thermometer. Radiative transitions from $^5D_0$ to the $^7F_0$, $^7F_1$, $^7F_2$, $^7F_3$, and $^7F_4$ levels produce emission bands at approximately 575 nm, 590 nm, 615 nm, 650 nm, and 700 nm, respectively. In the case of $LaGaO_3$ doped with $Eu^{3+}$, increasing the temperature induces a structural phase transition from the low-temperature (LT) orthorhombic phase with $C_s$ point symmetry to the high-temperature (HT) rhombohedral phase with $D_{2d}$ symmetry, occurring at approximately 450 K. This transition reduces the number of Stark levels for certain multiplets due to the increased symmetry. For instance, the Stark levels of $^7F_1$ and $^7F_2$ decrease from 3 and 5, respectively, in the LT phase to 2 and 3 in the HT phase. These changes are clearly reflected in the luminescence spectra recorded at 83 K and 550 K (Figure 2b) and are particularly evident in the normalized luminescence maps shown in Figure 2c-f. Although some Stark components overlap spectrally, a noticeable reduction in their number is observed with the phase transition. Additionally, the energies of the Stark levels shift, a feature that enhances thermometric utility by reducing spectral overlap between bands originating from different phases, thereby simplifying temperature readouts.

An essential characteristic of $Eu^{3+}$ ions for analyzing structural changes in $LaGaO_3$ is the nature of the $^5D_0 \rightarrow ^7F_2$ transition, which is electric dipole-induced and sensitive to local symmetry changes. In contrast, the $^5D_0 \rightarrow {}^7F_1$ transition is magnetic dipole-induced and largely unaffected by symmetry variations. Therefore, the intensity ratio ($LIR_1$) of these transitions, defined as:

$$LIR_1 = \frac{\int_{610nm}^{630nm} \left({}^5D_0 \rightarrow {}^7F_2\right) d\lambda}{\int_{580nm}^{600nm} \left({}^5D_0 \rightarrow {}^7F_1\right) d\lambda} \qquad (3)$$

can be employed to monitor structural changes in the material[58,59]. Using LaGaO$_3$:0.25%Eu$^{3+}$ as a reference sample, $LIR_1$ increases from 3.08 in the LT phase to 3.8 in the HT phase, indicating that symmetry changes enhance the intensity of the $^5D_0 \rightarrow {}^7F_2$ band relative to the $^5D_0 \rightarrow {}^7F_1$ band. An analysis of the effect of Eu$^{3+}$ ion concentration on the spectroscopic properties of LaGaO$_3$ reveals that while the shape and spectral position of bands unrelated to Eu$^{3+}$ remain unchanged (Figure 2 g), $LIR_1$ at room temperature monotonically increases from 3.18 for 0.1% Eu$^{3+}$ to 3.58 for 5% Eu$^{3+}$ (Figure 2 h). This suggests that higher Eu$^{3+}$ concentrations induce structural changes that favor the electric dipole transition. Several explanations for this phenomenon were considered. One possibility is reabsorption[81] where the $^5D_0 \rightarrow {}^7F_1$ band intensity decreases at higher Eu$^{3+}$ concentrations, enhancing the apparent intensity of the $^5D_0 \rightarrow {}^7F_2$ band. However, this effect is unlikely since the $^7F_1$ level is not the ground state, lowering the probability of reabsorption, and the band shape remains unaffected by dopant concentration. Another explanation involves changes in the phase transition temperature due to ionic radius mismatches between Eu$^{3+}$ and La$^{3+}$ ions. However, the smaller radius of Eu$^{3+}$ ions would theoretically increase the phase transition temperature, contradicting the observed trend[53]. The most plausible explanation is that local structural changes, such as unit cell contraction, occur with increased Eu$^{3+}$ doping, as confirmed earlier in this study. Interestingly, luminescence kinetics, characterized by an average decay time ($\tau_{avr}$) of approximately 1.2 ms, shows minimal sensitivity to $Eu^{3+}$ concentration, remaining nearly constant across all doping levels (Figure 2 i).

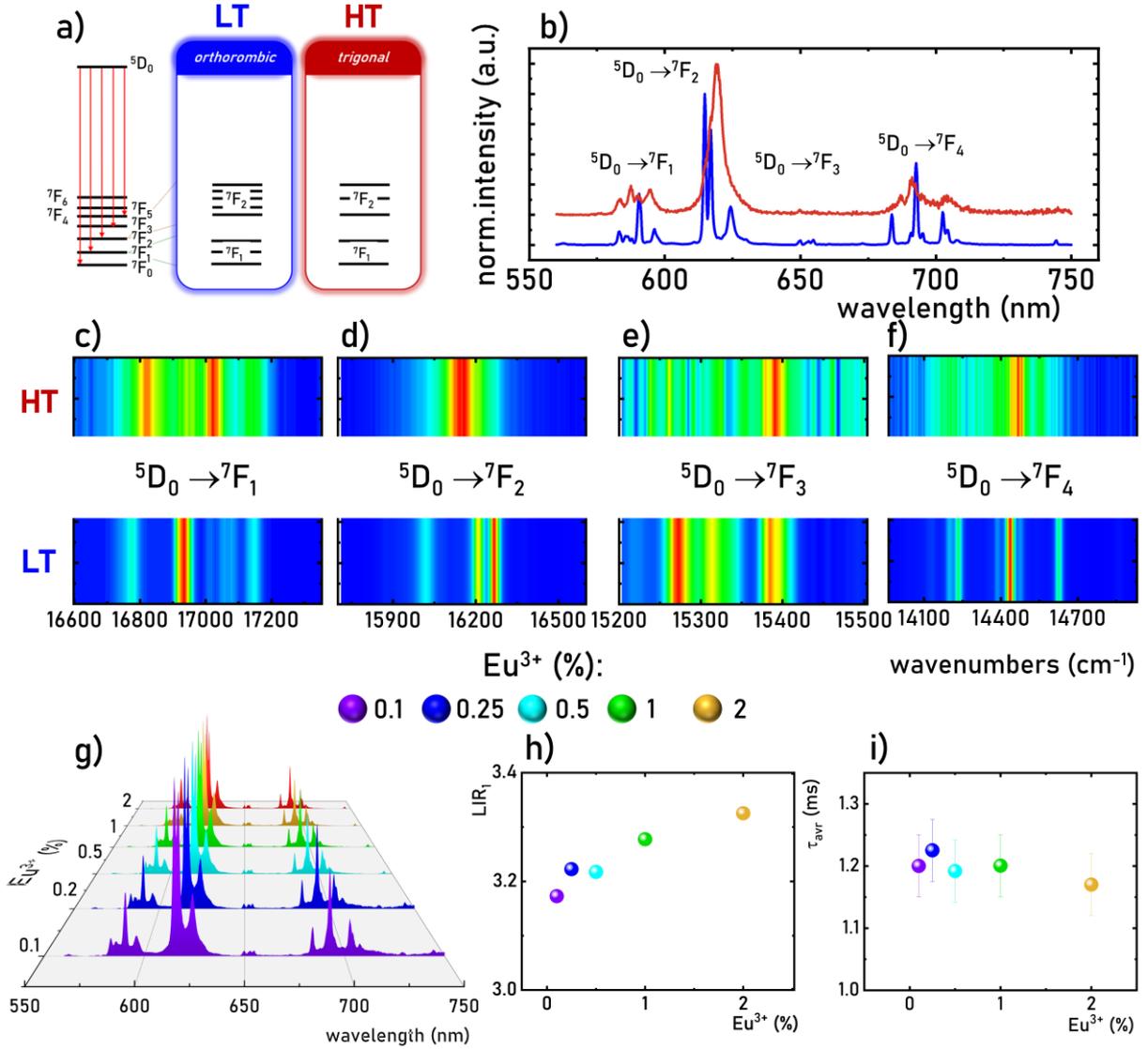

**Figure 2.** Simplified energy diagram of $Eu^{3+}$ ions with the Stark splitting of the $^7F_1$ and $^7F_2$ multiplets in the low temperature (LT) orthorhombic and high-temperature (HT) trigonal structures of $LaGaO_3:Eu^{3+}$-a) comparison of emission spectra of $LaGaO_3:0.25\%Eu^{3+}$ measured at 83 K (blue line) and 550 K (red line) which can be considered as a representative for LT and HT phases of $LaGaO_3:Eu^{3+}$ respectively ($\lambda_{exc}$=285nm)-b), normalized luminesce maps of LT and HT phases of $LaGaO_3:Eu^{3+}$ in the wavenumber ranges corresponding to the $^5D_0\rightarrow^7F_1$ -c), $^5D_0\rightarrow^7F_2$ – d), $^5D_0\rightarrow^7F_3$ – e) and $^5D_0\rightarrow^7F_4$ -f) emission bands of $Eu^{3+}$ ions; the influence of normalized room

temperature emission spectra of LaGaO$_3$:Eu$^{3+}$ for different concentration of Eu$^{3+}$ ions-f) the influence of Eu$^{3+}$ concentration on the *LIR$_1$* -h) and $\tau_{avr}$ ($\lambda_{exc}$=285nm, $\lambda_{em}$=615nm) – i) measured at room temperature.

Luminescence spectra of LaGaO$_3$:Eu$^{3+}$ measured as a function of temperature reveal that, with increasing temperature, not only does the luminescence intensity decrease, but the emission spectrum shape also changes, consistent with the mechanism described earlier (Figure 3a). Normalized luminescence spectra show a sharp decrease in the lines associated with Eu$^{3+}$ emission from the orthorhombic phase of LaGaO$_3$ and a concurrent increase in the lines corresponding to the emission from the trigonal phase at temperatures above approximately 440 K (Figure 3b and 3c). While these spectral changes are evident across all Eu$^{3+}$ emission bands, the increased number of Stark levels associated with higher J multiplets, such as those beyond $^7F_3$, can obscure the distinction between signals originating from the two phases. Consequently, the focus of this study is on the $^5D_0 \rightarrow {}^7F_1$ and $^5D_0 \rightarrow {}^7F_2$ transitions, where these changes are more easily discernible (Figure 3b and 3c, respectively). Interestingly, while individual Eu$^{3+}$ bands exhibit pronounced shape changes within the narrow temperature range corresponding to the phase transition, the intensity ratio *LIR$_1$* increases monotonically and gradually over the entire temperature range up to approximately 500 K, after which it begins to fluctuate (Figure 3d). A detailed analysis of the intensities of the limiting Stark lines, marked in Figures 3b and 3c, indicates that all lines decrease with increasing temperature due to thermal quenching of luminescence (Figure 3e). However, for spectral ranges associated with the LT phase of LaGaO$_3$:Eu$^{3+}$, a sharp intensity decrease is observed between 420 K and 500 K, coinciding with the phase transition, which was confirmed based on DSC studies for all samples (Figure 1d). This difference in the monotonic temperature dependence of the LT and HT phase intensities suggests its potential use in luminescence thermometry. To explore this further, the parameters *LIR$_2$* and *LIR$_3$* were defined as follows:

$$LIR_2 = \frac{Eu^{3+}(\text{rhombohedral})}{Eu^{3+}(\text{orthorombic})} = \frac{\int_{619nm}^{622nm} \left(^5D_0 \to {}^7F_2\right)d\lambda}{\int_{614nm}^{615nm} \left(^5D_0 \to {}^7F_2\right)d\lambda} \quad (4)$$

$$LIR_3 = \frac{Eu^{3+}(\text{rhombohedral})}{Eu^{3+}(\text{orthorombic})} = \frac{\int_{586nm}^{588nm} \left(^5D_0 \to {}^7F_2\right)d\lambda}{\int_{590nm}^{591nm} \left(^5D_0 \to {}^7F_2\right)d\lambda} \quad (5)$$

Both parameters exhibit analogous thermal behavior (Figure 3f). From 83 K to approximately 420 K, only slight changes in $LIR_2$ and $LIR_3$ values are observed. Above this temperature, a sharp increase is detected, with $LIR_3$ increasing by almost fourfold and $LIR_2$ increasing by nearly ninefold. Beyond 500 K, further temperature increases have little impact on $LIR_2$ and $LIR_3$ values until 600 K, above which a reduction in both parameters occurs due to the diminished luminescence intensity of $Eu^{3+}$ ions. The thermal variation of $LIR_2$ and $LIR_3$ can be quantitatively described using the relative sensitivity ($S_R$), a key thermometric parameter defined as:

$$S_R = \frac{1}{LIR}\frac{\Delta LIR}{\Delta T}\cdot 100\% \quad (6)$$

where $\Delta LIR$ shows the change of $LIR$ corresponding to change of temperature $\Delta T$. The thermal changes in $LIR_2$ and $LIR_3$ are reflected in their respective $S_R$ values, which reach maxima of $S_R=3.8\%\,K^{-1}$ for $LIR_3$ and $S_R=6\%\,K^{-1}$ for $LIR_2$ at approximately 480 K (Figure 3g).

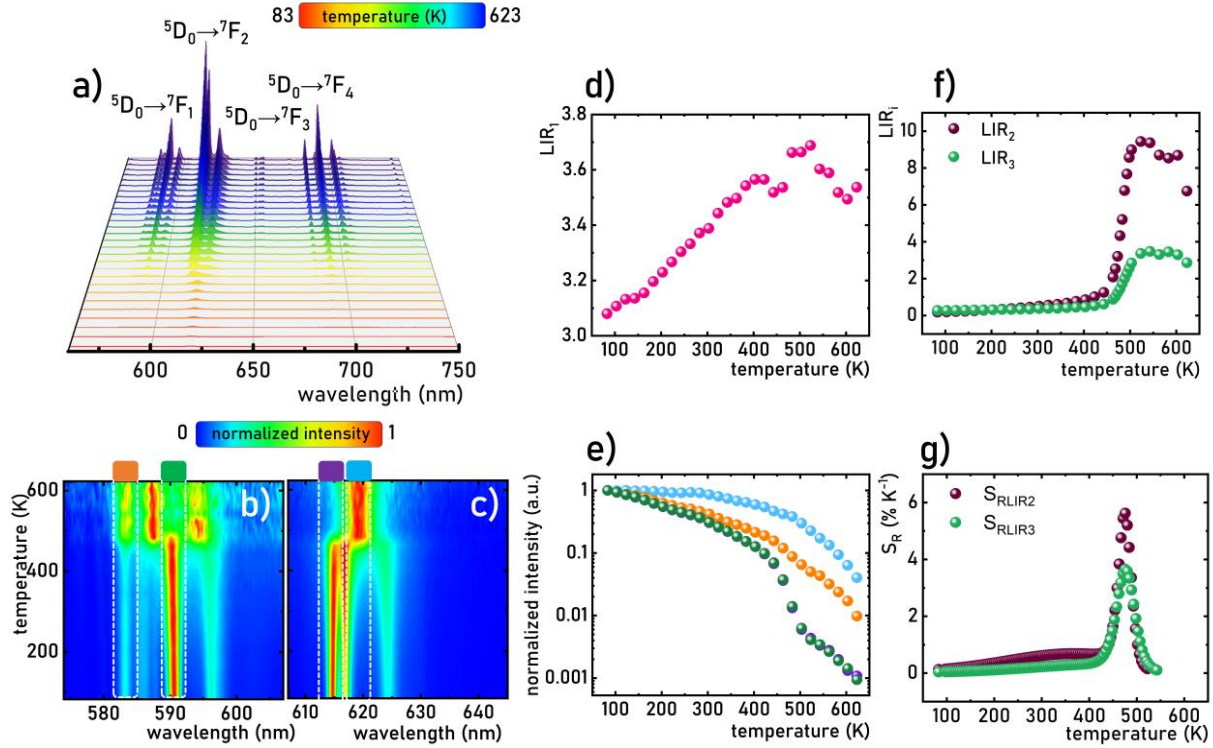

**Figure 3**. Emission spectra of LaGaO$_3$:0.25%Eu$^{3+}$ measured as a function of temperature -a); normalized luminescence thermal maps presented in the spectral range corresponding to the $^5D_0 \rightarrow {^7F_1}$ -b) and $^5D_0 \rightarrow {^7F_2}$ -c) electronic transitions; thermal dependence of *LIR$_1$* for the LaGaO$_3$:0.25%Eu$^{3+}$-d); thermal dependence of emission intensities of Eu$^{3+}$ integrated in the spectral ranges marked in Figure 3b and 3c – e), thermal dependence of *LIR$_2$* and *LIR$_3$* -f) and corresponding *S$_R$*-g).

Given that the highest relative sensitivity values were obtained for *LIR$_2$*, further analysis focuses on this parameter. It is well-established that altering the concentration of dopant ions with ionic radii differing from those of the host material ions typically results in a shift in the phase transition temperature. To explore this effect, the thermometric properties of LaGaO$_3$:Eu$^{3+}$ were studied as a function of Eu$^{3+}$ ion concentration. For all tested Eu$^{3+}$ concentrations, increasing temperature induced a similar *LIR$_2$* behavior (Figure 4a). Initially, *LIR$_2$* increased slightly with temperature until the phase transition temperature, at which point a sharp, several-fold increase was observed, followed by saturation. Importantly, the temperature at which this rapid increase occurred shifted to higher values with increasing Eu$^{3+}$ concentration, ranging from 420 K for 0.1% $Eu^{3+}$ to 500 K for 2% $Eu^{3+}$. Accordingly, the thermal dependence of $S_R$ also exhibited a shift in its maximum values toward higher temperatures with increasing $Eu^{3+}$ concentration (Figure 4b). Analyzing the maximum $S_R$ values for different $Eu^{3+}$ concentrations revealed a monotonic decrease from 6 %$K^{-1}$ for 0.1% $Eu^{3+}$ to 3% $K^{-1}$ for 2% $Eu^{3+}$ (Figure 4c). This phenomenon is commonly observed in materials undergoing phase transitions. Furthermore, the temperature corresponding to the maximum $S_R$ increased monotonically with $Eu^{3+}$ concentration, reflecting the shift in the phase transition temperature. This highly significant aspect underscores that the thermometric performance of $LaGaO_3$:$Eu^{3+}$ can be effectively tuned by adjusting the $Eu^{3+}$ concentration. Notably, this shift exceeds 100 K as the $Eu^{3+}$ concentration increases from 0.1% to 2% (Figure 4d). In addition to relative sensitivity, the temperature uncertainty ($\delta T$) is a critical parameter for evaluating the performance of a luminescent thermometer. The $\delta T$ can be determined using the following equation[37]:

$$\delta T = \frac{1}{S_R}\frac{\delta LIR}{LIR} \qquad (7)$$

The results indicate a strong correlation between high $S_R$ values and low $\delta T$. The lowest $\delta T$, approximately 0.2 K, was obtained for $LaGaO_3$:0.5%$Eu^{3+}$ (Figure 4e). As $S_R$ decreases with increasing $Eu^{3+}$ concentration, coupled with reduced emission intensity, $\delta T$ values increase for higher $Eu^{3+}$ concentrations.

Another crucial parameter from an application perspective is the thermal operating range of the luminescent thermometer. Although this range can be defined in various ways for monotonic thermometric parameters, in this work, it is considered as the range where $S_R$>0.5% $K^{-1}$. For $LaGaO_3$:0.1%$Eu^{3+}$, this range spans 200 K to 470 K, and it broadens with increasing $Eu^{3+}$ concentration, reaching 200 K to 610 K for $LaGaO_3$:2%$Eu^{3+}$. However, as the thermal operating range widens, $S_R$ decreases, and δT increases (Figure 4f). These findings demonstrate that by varying the $Eu^{3+}$ ion concentration, the parameters of the ratiometric luminescent thermometer based on $LaGaO_3:Eu^{3+}$ can be optimized to meet the specific requirements of different applications. This adaptability underscores the material's versatility and potential for practical use in a range of thermal sensing scenarios.

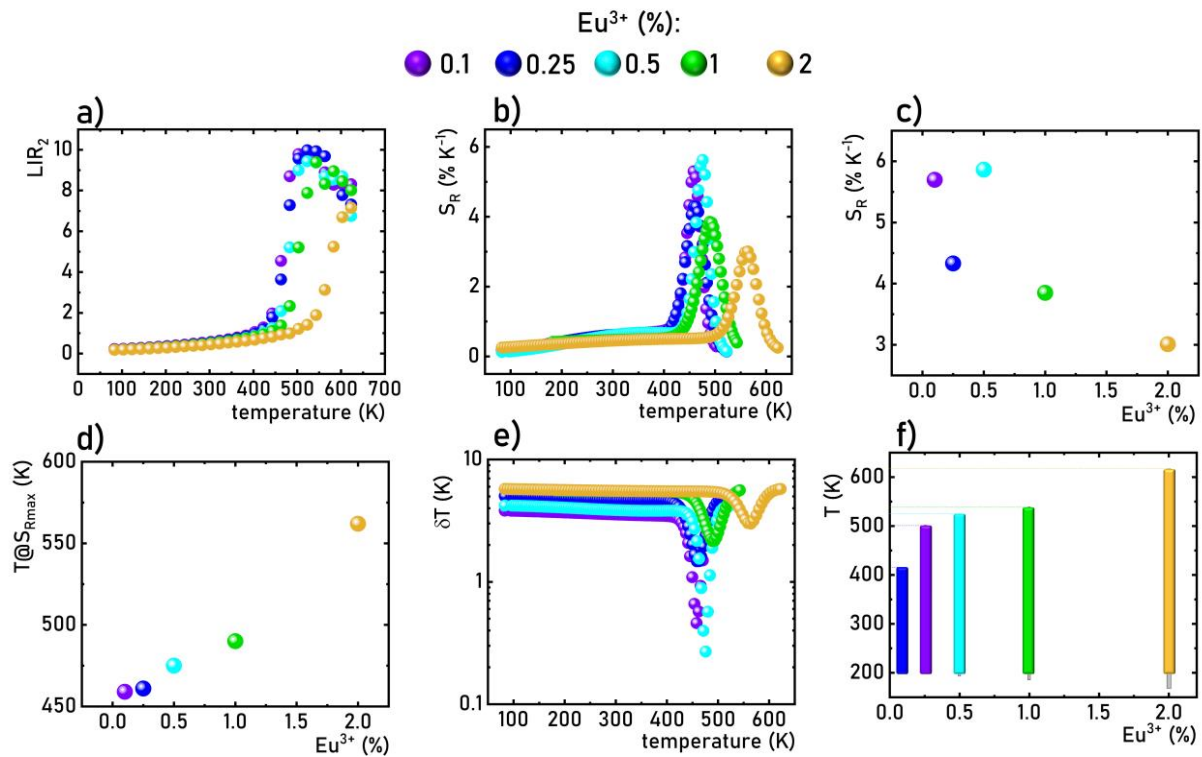

**Figure 4.** Thermal dependence of $LIR_2$ for different concentration of $Eu^{3+}$ ions for $LaGaO_3:Eu^{3+}$-a) and corresponding $S_R$-b); the influence of the $Eu^{3+}$ concentration of the maximal value of $S_R$ -c) and temperature at which $S_{Rmax}$ was observed-d), thermal dependence of $\delta T$-e), the influence of $Eu^{3+}$ concentration on the operating range of ratiometric luminescence thermometer based on $LIR_2$-f).

The use of thermally induced changes in the spectroscopic properties of phosphors resulting from structural phase transitions is a relatively novel concept, which explains the limited number of studies reported in the literature[51,53–56,67,69,70]. Most of these studies focus on LiYO$_2$, a material exhibiting a structural transition from a low-temperature monoclinic phase to a high-temperature tetrahedral phase. The phase transition temperature in LiYO$_2$ can be tuned over a range centered around 300 K[61,62,64] by modifying the chemical composition of the matrix, the concentration of dopant ions, and the material's morphology. Various luminescent ions, such as Eu$^{3+}$[53,67], Er$^{3+}$[54], and Pr$^{3+}$[68], have been utilized to exploit the significant structural changes in this material, enabling the development of thermometers operating in both the visible and infrared spectral ranges. The rapid changes in the spectroscopic properties of dopant ions within LiYO$_2$ have facilitated the achievement of exceptionally high sensitivity values, such as 23% K$^{-1}$ for LiYO$_2$:Pr$^{3+}$ [68] or even 35.24 % K$^{-1}$ for LiYO$_2$:Dy$^{3+}$ [56]. However, the thermal operating range of these thermometers is relatively narrow, typically oscillating around room temperature. Efforts to shift the operating range have involved co-doping with optically passive ions of ionic radii differing from the host material cation (Y$^{3+}$). For example, introducing smaller-radius ions has lowered the operating range to as low as 180 K (e.g., LiYO$_2$:1%Eu$^{3+}$, 30%Yb$^{3+}$)[54], while doping with larger-radius ions has extended the range to higher temperatures, such as 550 K (e.g., LiYO$_2$:1%Eu$^{3+}$, 40%Gd$^{3+}$). Despite these advancements, shifting the phase transition temperature to 550 K for LiYO$_2$:1%Eu$^{3+}$, 40%Gd$^{3+}$ results in a significant reduction in relative sensitivity, dropping to 1.4% K$^{-1}$, compared to 12% K$^{-1}$ for its Eu$^{3+}$-only counterpart. From this perspective, the sensitivity values achieved for LaGaO$_3$:Eu$^{3+}$, as described in this work, are particularly noteworthy. They demonstrate the potential to extend the thermal operating range of luminescence thermometers based on structural phase transitions to higher temperatures while maintaining high relative sensitivity values. This advancement is critical for expanding the applicability of phase transition-based luminescence thermometry in practical applications.

**Table 1**. Comparison of the thermometric performance of ratiometric luminescence thermometers based on the first-order structural phase transition

| Thermometer | LT phase | HT phase | LIR | $S_{Rmax}$ [% K$^{-1}$] | T@$S_{Rmax}$ [K] | Ref |
|---|---|---|---|---|---|---|
| LiYO$_2$:5%Yb$^{3+}$ | monoclinic | tetragonal | $^2F_{5/2} \to {}^2F_{7/2}$ / | 5.3 | 280 | 70 |

| | | | $^2F_{5/2} \to {}^2F_{7/2}$ | | | |
|---|---|---|---|---|---|---|
| LiYO$_2$:Pr$^{3+}$ | monoclinic | tetragonal | $^3P_0 \to {}^3H_4$ / $^1D_2 \to {}^3H_4$ | 23.04 | 329 | 68 |
| LiYO$_2$:1%Eu$^{3+}$ | monoclinic | tetragonal | $^5D_0 \to {}^7F_2$ / $^5D_0 \to {}^7F_2$ | 12.5 | 305 | 67 |
| LiYO$_2$:1%Eu$^{3+}$,30%Yb$^{3+}$ | monoclinic | tetragonal | $^5D_0 \to {}^7F_2$ / $^5D_0 \to {}^7F_2$ | 2.1 | 180 | 53 |
| LiYO$_2$:1%Eu$^{3+}$,40%Gd$^{3+}$ | monoclinic | tetragonal | $^5D_0 \to {}^7F_2$ / $^5D_0 \to {}^7F_2$ | 1.4 | 550 | 53 |
| LiYO$_2$:0.1%Nd$^{3+}$, | monoclinic | tetragonal | $^4F_{3/2} \to {}^4I_{9/2}$ / $^4F_{3/2} \to {}^4I_{9/2}$ | 7.9 | 291 | 52 |
| LiYO$_2$: 1%Er$^{3+}$, 10%Yb$^{3+}$ | monoclinic | tetragonal | $^4S_{3/2} \to {}^4I_{15/2}$ / $^4S_{3/2} \to {}^4I_{15/2}$ | 2.5 | 240 | 54 |
| LiYO$_2$: Dy$^{3+}$, | monoclinic | tetragonal | $^4F_{9/2} \to {}^6H_{13/2}$ / $^4F_{9/2} \to {}^6H_{13/2}$ | 35.24 | 310 | 56 |
| LaGaO$_3$:0.1%Eu$^{3+}$ | orthorhombic | trigonal | $^5D_0 \to {}^7F_2$ / $^5D_0 \to {}^7F_2$ | 5.4 | 456 | This work |
| LaGaO$_3$:2%Eu$^{3+}$ | orthorhombic | trigonal | $^5D_0 \to {}^7F_2$ / $^5D_0 \to {}^7F_2$ | 3 | 560 | This work |

**Conclusions**

In this study, the spectroscopic properties of LaGaO$_3$:Eu$^{3+}$ were investigated as a function of temperature to develop a luminescent thermometer based on its structural phase transition. Differential scanning calorimetry studies revealed that the structural phase transition from the low-temperature orthorhombic phase to the high-temperature trigonal phase occurs at temperatures ranging from 456 K for LaGaO$_3$:0.1%Eu$^{3+}$ to 570 K for LaGaO$_3$:2%Eu$^{3+}$, depending on the concentration of Eu$^{3+}$ ions. This transition is accompanied by structural changes in the local symmetry of La$^{3+}$ ions, which alters from $C_s$ to $D_3$ symmetry. These changes significantly affect the spectroscopic properties of Eu$^{3+}$ ions occupying these crystallographic positions, leading to variations in the intensity ratio of the $^5D_0 \to {}^7F_2$ band to the $^5D_0 \to {}^7F_1$ band and modifications in the number of Stark levels into which the $^7F_J$ multiplets are split. As a result, the intensity ratio of Stark lines corresponding to the rhombohedral and orthorhombic phases was employed as a thermometric parameter. This parameter exhibited a sharp increase around the phase transition temperature, achieving a maximum sensitivity of 5.4% K$^{-1}$ at 456 K for LaGaO$_3$:0.1%Eu$^{3+}$. An increase in the Eu$^{3+}$ concentration led to a monotonic decrease in $S_{Rmax}$ values, reaching 3% K$^{-1}$ at 560 K for LaGaO$_3$:2%Eu$^{3+}$. Additionally, higher Eu$^{3+}$ concentrations extended the thermal operating range, with the maximum thermal uncertainty remaining low at 0.2 K. The results clearly demonstrate that LaGaO$_3$ can serve as an excellent host material for luminescence thermometers based on structural phase transitions in the temperature range above 400 K. This study represents a significant step toward expanding the thermal range in which phase transition-based luminescent thermometers can operate, making this innovative approach increasingly viable for high-temperature applications.

## Acknowledgements

This work was supported by the National Science Center (NCN) Poland under project no. DEC-UMO-2022/45/B/ST5/01629.